\documentclass{aa}   
\usepackage{graphics,latexsym,amssymb,times,psfig}   
   
\def\gsim{ \lower .75ex \hbox{$\sim$} \llap{\raise .27ex \hbox{$>$}} }    
\def\lsim{ \lower .75ex\hbox{$\sim$} \llap{\raise .27ex \hbox{$<$}} }    
   
\begin{document}   
   
\title{Reconsidering the origin of the X--ray emission lines in GRB 011211}   
   
   
\author{Fabrizio Tavecchio \inst{1}, Gabriele Ghisellini \inst{1} and    
Davide Lazzati, \inst{2}} 
\offprints{F. Tavecchio; tavecchio@merate.mi.astro.it}   
  
\institute{  
INAF--Osserv. Astron. di Brera, via Bianchi 46, I--23807 Merate, Italy;   
\and   
Institute of Astronomy, University of Cambridge, Madingley Road, 
CB3 0HA Cambridge, UK}

\date{Received 2003}   
    
\titlerunning{Reflection model for GRB011211}   
\authorrunning{Tavecchio, Ghisellini \& Lazzati}   
   
\abstract{ We reanalyze the {\it XMM--Newton} data of GRB 011211
showing that the spectral features, interpreted by Reeves et
al. (2002, 2003) as due thermal emission from a collisionally ionized
plasma, can be also reproduced by a reflection model (with ionization
parameter $\xi\sim 10^2$).  We discuss the implications of this
interpretation, estimating the total mass required in the simplified
case of a funnel geometry.  We conclude that a moderate clumping of
the reprocessing material (corresponding to a filling factor of the
order of $f\sim 10^{-3}$) is required.  Finally we show that, if this
interpretation is correct, a bright quasi--thermal component is
expected in the optical--UV band (containing about $90\%$ of the
luminosity of the illuminating continuum), whose presence can be used
to test the reflection model.
}   
\maketitle   
   
\section{Introduction}   
   
The study of spectral features (absorption, emission lines) in the 
X--ray spectra of afterglows of Gamma-Ray Bursts is thought to be one 
of the most powerful tools to probe the circumburst environment.
In fact, even if the direct observation of the central engine 
is not possible, it might be possible to get indirect indication of 
the composition and structure of the GRB environment using emission 
lines (for recent reviews see Lazzati 2002a and B\"ottcher 
2002). Until now, emission features have been detected (in some cases 
with only marginal significance) in 7 GRBs, namely GRB 970508 (Piro et 
al. 1999), GRB 970828 (Yoshida et al. 2000), GRB 991216 (Piro et 
al. 2000), GRB 000214 (Antonelli et al. 2000), GRB 011211 (Reeves et 
al. 2002, 2003), GRB 020813 (Butler et al. 2003) and GRB 030227 
(Watson et al. 2003).  In most of these cases the X--ray spectrum 
shows the evidence of the Iron line, largely interpreted as due to 
recombination, either in a photoionized (e.g. Vietri et al. 2001) or 
in a collisionally--ionized (thermal) plasma (e.g. Paerels et 
al. 2000). 
   
Recently, Reeves et al. (2002, 2003) used a {\it thermal} model to fit 
the X--ray spectrum of the afterglow of GRB 011211 observed by {\it 
XMM--Newton} (unfortunately the observation is affected by pointing 
problems and the real significance of the lines has been criticized on 
different grounds by Borozdin \& Trudolyubov 2002 and Rutledge \& Sako 
2003). The residuals of a fit to the data with a pure absorbed 
power--law model show evident features located at 0.7, 0.9 and 1.2 
keV.  Reeves et al. show that a good fit to the data can be obtained 
assuming a thermal plasma model ($kT\sim 4 $ keV) enriched in light 
metals (Ca, Mg, Si, S, Ar).  The viability of such a model is 
critically discussed by Lazzati (2003) which concludes that a thermal 
origin of the emission puts severe constraints on the parameters of 
the emission region: in particular the emitting gas must be strongly 
clumped, with volume filling factors of the order of $\sim 
10^{-6}-10^{-7}$. Another problem comes from the fact that, in order 
to fit the spectrum with the thermal VMEKAL model, Reeves et al. are 
forced to assume overabundance of Mg, S, Si, Ar, Ca (about a factor of 
10 larger than the solar abundance), but a solar abundance of iron, 
clearly not easily explainable in a standard hypernova scenario 
(Woosley 1993), since the exploding star is thought to produce a large 
amount of Ni, Co and Fe (e.g. Woosley \& Weaver 1995). 
   
Reeves et al. (2003) also considered the possibility that the 
spectrum is produced through the reflection by an ionized slab (Ross 
\& Fabian 1993; Ballantyne \& Ramirez-Ruiz 2001), but concluded that  
this model (either ``pure'' or with an underlying power--law) cannot  
reproduce the data. However (as explicitly noted by the authors), most  
of the discrepancy between the model and the data is localized in the  
region around 1 keV, where lines from elements {\it not included} in  
the ionized reflection model, such as S, Ar and Ca, are expected.  
 
More recently, Butler et al. (2003) and Watson et al. (2003) 
reported the possible detection of emission lines of light elements 
(but not Iron) in the {\it Chandra} HETGS spectrum of GRB 020813 and 
in the {\it XMM-Newton} spectrum of the afterglow of GRB 030227. These 
features can be identified as lines of hydrogen--like SXVI and SiXIV 
(GRB 020813) and from helium/hydrogen--like Mg, Si, S, Ar and Ca (GRB 
030227). Also in these cases a thermal model could fit the data, but 
the authors noted that these lines can also originate from reflection 
by material illuminated by the burst/afterglow (even though specific 
fits for this case are not presented for GRB 020813). Even more 
interestingly, in the case of GRB 030227 the lines are not detected in 
the first part of the observation (started about 11 h after the burst) 
but appear only in the last 10 ksec, approximately 20 h after the  
trigger. 
 
All these lines are detected at the $\sim 3\sigma$ level: although we
are waiting for a more secure detection, casting away any doubt about
their reality, the information carried by their presence are so
important to justify their study.  We therefore assume that these
lines are real, and in this paper we reanalyze the {\it XMM--Newton}
data of GRB 011211 to show that it is possible to reproduce them
modelling the spectrum with the reflection model, once lines from
light elements are properly considered.  We discuss the case in which
the reflector has a funnel geometry and we propose simple
observational tests for this scenario that could be performed with
future missions such as {\it SWIFT}.  We find a consistent scenario
able to account for the observations of the emission lines of light
metals (but no iron) in GRB 011211, and we point out that if this
scenario is correct there should be an intense optical--UV--soft
X--ray emission accompanying the emergence of the emission lines.
This is therefore a strong test for the reflection interpretation of
the X--ray emission lines.
 
The plane of the paper is the following: in Sect. 2 we report the results 
of the fit with the reflection model. In Sect. 3 we derive and discuss the 
constraints on the GRB environment and the properties of the   
reflection model. 
 
In the paper the following cosmology is assumed: $H_0=70$ km s$^{-1}$   
Mpc$^{-1}$ $\Omega_\Lambda=0.7$ and $\Omega=1$.

\section{Spectral fits}   
   
{\it XMM--Newton} data of GRB 011211 processed through the standard   
pipeline are publicly available 
\footnote{ ${\rm http://xmm.vilspa.esa.es/external/xmm\_sched/too}$}. 
We extract the PN spectrum of the first 5 ks, where lines are visible 
(Reeves et al. 2002, 2003), using the standard package SAS.  We fitted the 
spectrum using XSPEC (v.11).   
 
\subsection{Pure reflection} 
 
Unfortunately, the reflection model available in XSPEC (available as 
table model file; see the description of the model in Ballantyne, 
Iwasawa \& Fabian 2001) is calculated for two fixed values of the 
abundance, solar and 2 x solar. Moreover, as anticipated in the 
Introduction, it does not include light metals such as S, Ar and Ca. 
The reflection model assumes that a slab is illuminated by a power law 
continuum and the basic fitting parameter is the ionization parameter, 
defined as $\xi=L_{\rm ill}/(n R^2)$, where $L_{\rm ill}$ is the 
luminosity of the continuum, $n$ is the density of the illuminated 
slab and $R$ is the distance between the slab and the illuminating 
source. 
 
To include the contribution of light metals we fit the data using the 
following procedure: we start fitting the data by using the (absorbed 
with Galactic $N_{\rm H}=4.2\times 10^{20}$ cm$^{-2}$, Reeves et 
al. 2003) reflection model with free redshift.  In the fitting 
procedure we excluded all the bins below 0.3 keV, due to known 
problems with the calibration of the response matrix (e.g. Brinkmann 
et al. 2001).  As expected, the residuals of the best fit (which 
converges at $z=1.9$) show clear deviations around 0.7, 0.9, 1.1, 1.2 
and 1.5 keV.  This agrees with the analysis by Reeves et al. 
The line at 0.7 keV can be attributed to SiXIV.  The model includes 
this element, but the fit of the line is poor, since the data requires 
a brighter line (which may correspond to a larger abundance).  The 
other features can be identified as emission lines from elements not 
included in the reflection model (SXVI at 0.9 keV, ArXVII and SXV at 
1.1 keV, Ar XVIII at 1.2 keV and CaXX at 1.5 keV).  We then added 
Gaussian emission lines to account for these residuals.  The folded 
spectrum and the model/data ratio are shown in Fig. 1.  The unfolded 
spectrum is shown on Fig. 2.  The best fit ($\chi^2/d.o.f.=27.45/26$) 
parameters are: $Log \xi =2.25 \pm 0.30$, $z=1.93 \pm 0.10$ with an 
incident power--law of photon spectral index $\Gamma =2.43 \pm 0.20$. 
\begin{figure}   
\psfig{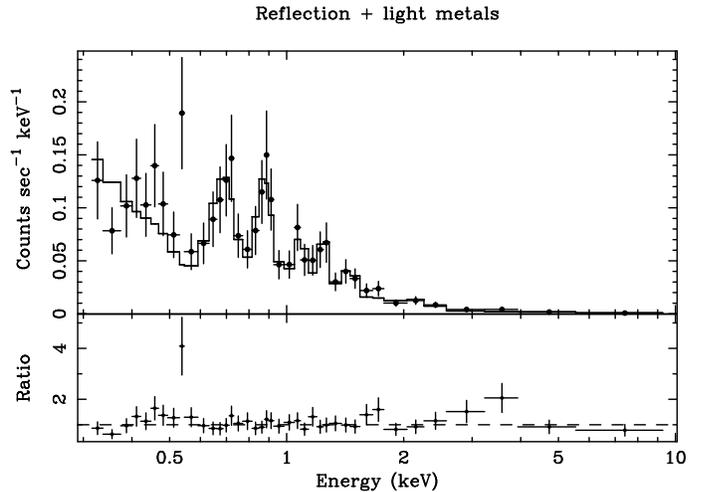}   
\caption{Spectrum (top panel) and data/model ratio (bottom panel) of 
the first 5 ksec of the PN data fitted with the reflection model and 
lines from light elements (see discussion in the text).} 
\end{figure}   
\begin{figure}   
\psfig{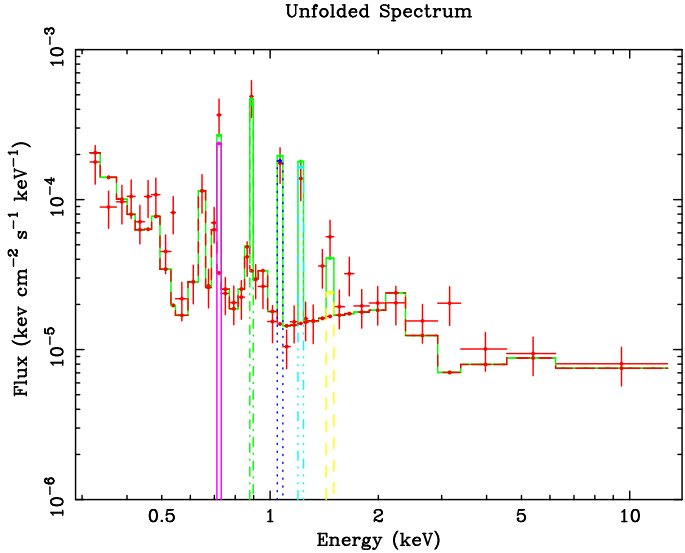}   
\caption{Unfolded spectrum of the PN spectrum fitted with the 
reflection model. The narrow lines at 0.7, 0.9, 1.1, 1.2, and 1,5 keV 
are the lines of the elements reported in Table 1, not included in the 
reflection model.} 
\end{figure}  
 
The parameters of the lines we have added, but that are not included
in the fitting model, are reported in Table 1.  Similarly to Reeves et
al. (2003) we find that the lines require a redshift $z=1.8-1.9$,
lower than the redshift of the GRB, $z=2.140\pm 0.001$ (Holland et
al. 2002), suggesting that the emitting material is moving with a
velocity $v\sim 0.1 c$. The luminosity of the Fe line, included in the
model is $2.3\times 10^{44}$ erg s$^{-1}$. For the possible presence
of the H-like Nickel (rest frame energy $E=8.102$ keV) we obtained an
upper limit to the equivalent width of $EW < 450$ eV (Reeves et
al. 2003 found $EW < 800$ eV).  The expected luminosities of light
elements, compared to iron, have been calculated by Lazzati et
al. (2002b).  The maximum contrast between the luminosity of the lines
of the light metals and the iron line is reached around $\xi \sim
10^2$.  For $\xi \sim 2\times 10^2$, required by our fit, we expect a
luminosity ratio of the order of 3, in agreement with the luminosities
reported in Table 1. Looking in detail at each element we note
that lines of S and Ar appear to be more luminous than expected, by a
factor of 2 and 5, respectively (note however that, due to the
relatively large uncertainties, the normalizations are consistent with
solar abundance within 1$\sigma $ and 2$\sigma $, respectively). This
discrepancy (present also in the fits presented by Reeves at al.)
could be ascribed to a larger abundance of these elements with respect
to the solar value.

Thus we can conclude that, once light metals are 
included, the reflection model provides a consistent fit of the {\it 
XMM--Newton} spectrum. 
 
\begin{table}   
\begin{center}   
\begin{tabular}{lllll}   
\hline\\    
Ion &$E_{\rm obs}$ & $E_{\rm rest}$ & $z$ & $L$ \\    
 & keV & keV &  & $10^{45}$ erg s$^{-1}$ \\   
\hline    
SiXIV& 0.71& 1.99 & 1.82 & 0.71 \\   
SXVI & 0.88& 2.60& 1.94 & 1.2\\   
ArXVII (SXV)& 1.08&  3.16 (3.14)& 1.93 (1.91) & 0.95 \\   
ArXVIII& 1.23& 3.30& 1.73& 0.95\\   
CaXX& 1.50& 4.07& 1.71&0.23\\   
\hline   
\end{tabular}   
\caption{Parameters of the lines added to the reflection model (see 
text for details). Luminosities refer to the pure reflection case; in 
the reflection+power--law case they have to scaled by a factor 0.7.} 
\end{center}   
\end{table}

\subsection{Reflection plus underlying power--law} 
 
Although a pure reflection model can account for the observed X-ray 
spectrum, it is possible, and even likely, that part of the flux 
originates from the afterglow continuum. It is important to stress 
that, unlike the case of AGNs, the continuum reprocessed through the 
reflection and that directly observed from the afterglow, 
simultaneously with the lines, {\it could be different}. In fact the 
radiation reprocessed by the reflector originates from the unobserved, 
rapidly changing burst/early--afterglow phase (hereafter illuminating 
continuum), while the afterglow radiation (observed when lines are 
detected) is produced much later (hereafter late afterglow continuum). 
 
We performed a series of fits to asses how much of the total emission 
can be ascribed to an underlying power--law continuum produced by the 
late afterglow continuum .  We fit the data with a model including a 
power--law and the reflection spectrum. We fix the relative 
normalization of the reflection spectrum and of the gaussian lines to 
the value found in the fit with the pure reflection case. 
 
First we constrain the two slopes, that of the illuminating continuum 
and that of the late afterglow power-law, to the same value. The fit 
converges to $\Gamma =2.26\pm0.1$, $Log \, \xi = 1.9 \pm 0.1$ with 
$\chi^2/d.o.f. = 23.71/31$.  The significantly improves with respect 
to the pure reflection case. The presence of a direct power-law is 
therefore suggested by the data. 
 
We then allowed the two photon indices to vary independently. In this 
case the best fit parameters are: $\Gamma _{\rm pl} 
=2.04^{+0.29}_{-0.08}$, $\Gamma _{\rm refl} =2.39^{+0.1}_{-0.06}$, $Log 
\, \xi = 2.1 \pm 0.1$ with $\chi^2/d.o.f. = 23.13/30$.  Within 90\% 
conf. level, the two slopes are consistent. Reflection contributes for 
$\sim $70\% of the flux. Correspondingly, the luminosity of the 
emission lines given in Tab.1 decreses by a factor 0.7.

\section{Discussion}   
   
We showed that the {\it XMM--Newton} PN data of GRB 011211 can be
satisfactorily fitted by the pure reflection model, if lines from the
elements not included in the spectral model are properly
considered. The ionization parameter $\xi$ is well constrained to be
of the order of $10^2$, since the data do not present a prominent iron
line while lines of light metals are luminous.  As mentioned, the
absence of a prominent iron line is naturally predicted when the
ionization parameter $\xi $ is around $10^2$ (Lazzati et
al. 2002b). In this case, due to the importance of the Auger effect,
the emission from iron is strongly depressed compared to the other
metals.
 
The reflection model can explain very different spectra, since 
variations of $\xi $ in different bursts could produce different 
reflection spectra, with prominent or almost absent lines 
(e.g. Ballantyne \& Ramirez-Ruiz 2001). Low values of the ionization 
parameter ($\xi <100 $) will produce a reflection component with 
evident lines.  Large ionization parameter ($\xi >10^4$), on the other 
hand, would produce almost featureless spectra, with a shape similar 
to the shape of the illuminating continuum.  For a narrow range of 
$\xi$ centered around $10^2$ the spectra will show luminous lines from 
light metals and a depressed K$_\alpha$ iron line. It may be argued 
that in the case of GRBs a well defined ionization parameter $\xi$ can 
hardly be attained. In the geometry we consider here the observer 
receives a spectrum resulting from the convolution of reflected 
components with different illuminating spectrum. Regions closer to the 
line of sight will contribute low $\xi$ components, while regions away 
from the line of sight will contribute higher $\xi$ components, since 
the time-delay is longer and the ionizing continuum is a monotonically 
decreasing function (at least in the afterglow phase). While this is 
true, it is clear that one of the components (and hence a well defined 
$\xi$) will dominate. Let us now consider the early X-ray afterglow as 
illuminator, since the GRB proper will have a too high $\xi$ to 
produce lines, as discussed above. If the illuminator has a power-law 
decay in time $L\propto{}t^{-\alpha}$, the more (less) luminous phase 
will dominate the observed reflected component for $\alpha>1$ 
($\alpha<1$). Unfortunately the X-ray afterglow has never been 
observed early enough to constrain its decay in the early phases, but 
it is not unreasonable to assume that the X-ray band lies initially 
below the cooling frequency, and that its decay slope is initially the 
same as the optical slope at later stages. In this case, bursts with 
an initially shallow decay in the optical, such as GRB~011211 (Holland 
et al. 2002) and GRB~020813 (Covino et al. 2003), should show 
low-ionization features, while bursts with initial fast decay such 
GRB~991216 (Halpern et al. 2000) should have X-ray spectra with strong 
iron lines. Even though it is not possible to make statistics with 
such a handful of cases, it is tantalizing that such a correlation is 
indeed respected in all cases. 
 
In the following we derive the geometrical and physical setup of the
reflecting material. We note that the actual geometry of the
funnel is unimportant for most of the estimates reported below, aimed
to find some basic quantities such as the overall size and the total
mass of the line emitting material.  The emitting layer is supposed to
be illuminated by an intense X--ray flux: this requires that either
the illuminating source is less collimated than the funnel (if it has
a conical profile), or that the funnel deviates from a perfect conical
geometry, being e.g.  parabolic.  In these conditions the funnel walls
can receive a non--negligible fraction of the ionizing flux.  We then
consider more complex geometries, involving clumps of overdense
material, possibly located close to the funnel surface, as illustrated
in Fig. \ref{funnel}.
  
\subsection{General constraints}   
   
Once the ionization parameter and the line luminosities are known, we   
can infer important clues on the physical state of the reflector. 
The ``observed" or fixed quantities are the following:  
\begin{enumerate} 
\item 
the luminosity of the line $L_{\rm line}$; 
\item 
its duration $t_{\rm line}$; 
\item 
the time $t_{\rm app}$ at which the line starts to be visible with the 
luminosity $L_{\rm line}$; 
\item 
the ionization parameter $\xi$;  
\item 
the efficiency $\eta$ of converting the illuminating continuum into the 
specific line; 
\item 
the total mass $M_{\rm T}$ (which we require not to exceed $\sim$10$M_\odot$). 
\end{enumerate} 
Note that the time at which lines appear ($t_{\rm app}$) is an
observable in principle, but it may require early observations of the
X--ray afterglow to be detected.  Indeed, it has been already
determined only in one case, (GRB 030227) while we have only an upper
limit for all other cases.  However, it is likely that with a short
pointing time, like the one provided by {\it SWIFT}, the number of
cases in which this parameter will be observable will strongly
increase.  Therefore, in view of a general discussion, we suppose that
$t_{\rm app}$ is known. For the specific case of GRB 011211 we only
have an upper limit to $t_{\rm app}$, since the lines were detected in
the first part of the observation.
 
Unknown physical quantities of the system are:  
\begin{enumerate} 
\item 
the distance $R$ between the illuminator and the line emitting material.  
\item 
the collimation angle $\theta$ of the illuminator (see Fig. 3). 
\item 
the electron density $n_{\rm e}$ (which, for the large ionization 
conditions discussed here is close to the proton density $n_p$); 
\item 
the illuminating energy $E_{\rm ill}$ (or, alternatively, the
illuminating luminosity $L_{\rm ill}$) directly intercepted by the
line-emitting material. Part of this illuminating energy can come
from the burst emission, and part from the early afterglow.  It is a
rapidly variable quantity, and for the sake of simplicity we will
assume that $L_{\rm ill}$ is an appropriate average over $t_{\rm
ill}$;
\item 
the time $t_{\rm ill}$ for which the illumination lasts; 
\item 
the number of lines photons, $k$, produced by each atom during $t_{\rm ill}$; 
\item 
the filling factor $f$; 
\item 
the height of the shell $\Delta R$;  
\end{enumerate} 
 
Thus we have a total of 6 quantities provided by observations and 8
unknown quantities.  Therefore we will consider $\theta$ and $t_{\rm
ill}$ as free parameters, even if their values are in any case limited
(for instance, ``reasonable" aperture angles of the illuminator should
be between 10 and 60 degrees; the illuminating time should be shorter
than the time for which the line is visible, $t_{\rm ill}< t_{\rm
line}$).
 
\begin{figure}   
\vspace{-2.4 truecm} 
\hspace{-1.5 truecm}
\psfig{figure=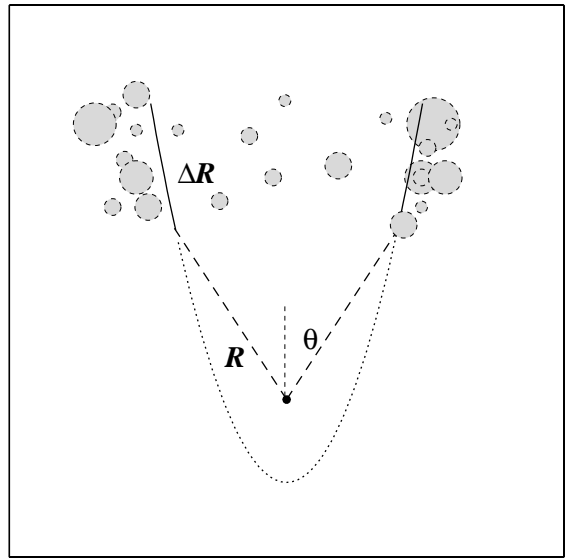,angle=0,width=11.5cm}   
\vspace{-4.0 truecm} 
\caption{Sketch of the geometry assumed for the reflecting material
(not to scale). The material is organized in clumps along the walls of
the funnel (assumed to have a parabolic shape). Clumps could also lie
along the line of sight.}
\label{funnel} 
\end{figure}   
 
In the following we outline how we can connect these two sets of 
parameters and how we can derive the parameters of the system by the 
observational quantities. All the time intervals ($t_{\rm ill}$, $ 
t_{\rm line}$, $ t_{\rm app}$) are intended as measured in the frame 
of the burst. 
We adopt the notation $Q \equiv 10^x Q_x$ and use cgs units. 
 
The size of the reflector is related to the time at which lines become 
visible: 
\begin{equation} 
R\, =\, {ct_{\rm app}\over 1-\cos \theta } 
\label{r} 
\end{equation} 
\noindent 
The total mass of the element emitting a specific line (denoted by the subscript $i$)  
is related to the number $N_i$ of photons in that line by: 
\begin{equation} 
M_i\, =\, {N_i W_i m_{\rm p} \over k} 
\label{mi} 
\end{equation} 
\noindent 
where $W_i$ is the atomic weight of the element. $k$ is the number 
of line photons a single atom can emit during the illumination phase 
(Lazzati et al. 1999): 
\begin{equation} 
k\, =\, {t_{\rm ill} \over t_{\rm rec}} 
\label{kdef} 
\end{equation} 
\noindent 
where the recombination time is: 
\begin{equation} 
t_{\rm rec}\, = \, \frac{1} {\alpha _{\rm r} n_{\rm e} } \,\,\, {\rm s}
\label{trec} 
\end{equation} 
\noindent 
$n_{\rm e}$ is the numerical density of the free electrons 
and $\alpha _{\rm r}$ is: 
\begin{equation} 
\alpha _{\rm r}=5.2\times 10^{-14} Z \lambda ^{1/2} \left[ 0.429 
+0.5 \ln(\lambda) + \frac{0.496}{\lambda ^{1/3}} \right] 
\label{alphar} 
\end{equation} 
\noindent 
where $\lambda =1.58\times 10^5 Z^2 T^{-1}$, $Z$ the atomic number of
the element and $T$ the electron temperature.
 
The total number of photons produced by atoms of the element $i$ can 
be written as: 
\begin{equation} 
N_i\, =\, {E_{\rm line}\over \epsilon_i}\, =\,   
{L_{\rm line} t_{\rm line} \over \epsilon _i} 
\label{nphot} 
\end{equation} 
\noindent 
where $E_{\rm line} = L_{\rm line} t_{\rm line}$ is the total energy in the 
line and $\epsilon_i$ is the energy of the line.  
The mass of the layer emitting the observed line is  
(using the mass abundance $A_i$ of the emitting element): 
%
%
%
\begin{equation} 
M_{\rm l}\, =\, {M_i \over A_i }\, =\,  
{L_{\rm line} t_{\rm line} W_i m_{\rm p}\over  k A_i \epsilon_i } 
\label{m1} 
\end{equation} 
\noindent 
This mass is contained into a layer of volume $V$, 
of thickness $\Delta x$, radius $R\sin\theta$ and height $\Delta R$. 
The material reflecting the illuminating continuum and 
producing the line is contained in a layer of height $\Delta x$ of 
unit scattering optical depth:  
$\tau _{\rm T}=\sigma _{\rm T}n_{\rm e} \Delta x =1$. 
Since the matter is highly ionized the density $n_{\rm e}$ 
of free electrons is the same as the proton density $n_{\rm p}$. 
This allows us to derive the mass of this layer in a way independent 
from Eq. (\ref{m1}) and also independent of the electron density:  
\begin{eqnarray} 
M_{\rm l}\,  
& =&\, n_{\rm p} m_{\rm p} V\, = \, 2\pi R \sin\theta \Delta R  
       \Delta x n_{\rm e} m_{\rm p}\, \nonumber \\ 
& =&\, {2\pi \over \sigma_{\rm T}} R \Delta R \sin \theta m_{\rm p}\, \nonumber \\ 
& \simeq& \,  5\pi R \Delta R \sin \theta  \,\,\, {\rm g}
\label{m2} 
\end{eqnarray} 
\noindent 
We then have two {\it independent} estimates of the mass involved in  
the line production, Eq. (\ref{m1}) and Eq. (\ref{m2}). 
By equating them we can derive the value of $k$ (which becomes  
independent of $n_{\rm e}$)  
\begin{equation} 
k\, =\, {W_i m_{\rm p} \over A_i \epsilon _i} \,  
{L_{\rm line} t_{\rm line} \over 5\pi R\Delta R \sin\theta} 
\label{k} 
\end{equation} 
\noindent From the definition  
(\ref{kdef}) and Eq. (\ref{trec}) and inserting $k$ from (\ref{k})  
we can derive the density: 
\begin{eqnarray} 
n_{\rm e}\, &=&\, {k  \over \alpha _{\rm r} t_{\rm ill}} \nonumber \\ 
             &=&\,  
{W_i m_{\rm p} \over A_i \epsilon _i}  
{L_{\rm line} t_{\rm line} \over 5\pi R\Delta R \sin \theta \alpha _{\rm r} t_{\rm ill}} \,\,\, {\rm cm}^{-3}
\label{n} 
\end{eqnarray} 
The ionization parameter is defined as: 
\begin{equation} 
\xi\, =\, {L_{\rm ill} \over n_{\rm e} R^2} 
\label{xidef} 
\end{equation} 
Inserting the density given by Eq. (\ref{n}), we have: 
\begin{equation} 
L_{\rm ill}\, =\, {1 \over 5\pi} \,  \xi \, {R\over \Delta R}\, 
{ W_i m_{\rm p}  E_{\rm line} \over 
A_i \epsilon_i \sin\theta \alpha _{\rm r} t_{\rm ill}} 
\,\, {\rm erg\, s^{-1}} 
\label{lill} 
\end{equation} 
The efficiency $\eta$ in producing the line is defined as: 
\begin{equation} 
\eta\, =\, {E_{\rm line} \over E_{\rm ill}}\, =\, 
{L_{\rm line} t_{\rm line}  \over L_{\rm ill} t_{\rm ill}} 
\label{etadef} 
\end{equation} 
Using Eq. (\ref{lill}) and Eq. (\ref{etadef}) we derive: 
\begin{equation} 
{\Delta R\over R}\, =\,  {1 \over 5\pi} \, \xi\,  \eta  \,  
{ W_i m_{\rm p}   \over 
 A_i \epsilon_i \sin\theta \alpha _{\rm r}} 
\label{drr} 
\end{equation} 
From Eq. (\ref{m2}) the mass of the layer turns out to be: 
\begin{equation} 
M_{\rm l}\, = \, R^2 \xi\, \eta \, {W_i m_p \over A_i \epsilon_i \alpha _{\rm r}} \quad {\rm g} 
\label{mlayer} 
\end{equation} 
The {\it total} amount of mass $M_*$ (i.e. contained in a shell of  
surface $4\pi R^2$ and height $\Delta R$ and assuming the same density)  
will be: 
\begin{equation} 
M_*\, =\, n_{\rm p}m_{\rm p} V_*\, =\, 4\pi R^3  
{\Delta R \over R} n_{\rm e} m_{\rm p} 
\label{mtot} 
\end{equation} 
\noindent 
We impose that this mass cannot exceed $M_T\sim 10 M_{\odot}$. If
$M_*$ exceeds this value, we assume that the material is
distributed inhomogeneously, more dense in the walls of the funnel or
organized in single clumps (see below). We parametrize the deviation
from the homogenoeus case by the filling factor:
\begin{equation} 
f\, =\, {M_T \over M_*} 
\label{clump} 
\end{equation} 
Note that there is in principle another equation relating $\Delta R$  
with the duration of the line $t_{\rm line}$ given by 
\begin{equation} 
\Delta R \, \sim \, { c t_{\rm line} \over \cos\theta}  
\label{dr} 
\end{equation} 
This equation is valid only in the case in which the duration of
the emission of the line is dominated by geometrical effects,
i.e. $t_{\rm ill} < \Delta R/c$. In principle, in this case it is
possible to fix the aperture angle of the funnel, using
Eq. (\ref{drr}) together with Eq. (\ref{dr}) and Eq. (\ref{r}):
\begin{equation} 
\tan\theta(1-\cos\theta) \,  
\sim \, {1 \over 5\pi} \, \xi \, \eta 
{W_i m_p \over A_i \epsilon_i \alpha _{\rm r}}\, {t_{\rm app} \over t_{\rm line}} 
\end{equation} 
\label{theta} 
For the application to the specific case of GRB 011211 we have 
preferred to use the angle $\theta $ as a free parameter, using 
Eq. (\ref{dr}) as a consistency check.  We do that because of the 
uncertainties related to the specific geometry. 

We stress that the above estimates have been derived for a
specific geometry, i.e. a uniform funnel over-dense with respect to
the rest of the remnant.  It is conceivable that the real situation is
much more complex: for instance, dense clumps of matter could be
embedded in a less dense medium, and be located also in the interior
of the funnel, not only on its walls (Fig. \ref{funnel}).  On the
other hand the values of the parameters we derive will not dramatically
change in this case, and can therefore be thought as indicative
numbers.

\subsection{Application to GRB 011211} 

As an illustrative example of the application of the estimates given
above we can apply our treatment to the specific case of GRB 011211.
For the observational quantities we set:
\begin{itemize} 
\item 
$L_{\rm line}=7\times 10^{44}$ erg s$^{-1}$ for the SiXIV line; 
\item 
$t_{\rm line}=5\times 10^3/(1+z)\sim 1.7\times 10^3$ seconds; 
\item 
$t_{\rm app} \le 4\times 10^4/(1+z)\sim 1.3\times 10^4$ seconds; 
\item 
$\xi=10^2\xi_2$;  
\item 
$\eta=10^{-2}\eta_{-2}$ (see Ghisellini et al. 2002); 
\item 
$M_{\rm T}=10M_\odot$ 
\end{itemize} 
 
In addition, we have used the redshift $z=2.14$; the line energy
$\epsilon_{SiXIV}=1.99$ keV and an abundance $A_i$ equal to twice the
solar value. The temperature of a photoionized plasma
illuminated with $\xi \sim 100$ is predicted to be in the range
$10^5-10^6$ K (e.g. Kallman \& McCray 1982).  We have then assumed
that the line emitting material is at $T=5\times 10^5$ K.  A variation
of the temperature, directly affecting the recombination rate, implies
a different amount of matter necessary to account for the emission and
therefore the estimate of the density and of the clumping factor.
Eq.(\ref{alphar}) indicates that the dependence of $t_{\rm rec}$ on
the temperature is $t_{\rm rec} \sim T^{3/4}$. A larger (lower)
temperature would decrease (increase) the recombination rate,
resulting in a larger (lower) clumping $f\sim T^{-3/4}$.

As discussed in Sect. 3.1, in the specific case of GRB
011211, since the appearence time of the line is unknown, the value of
$t_{\rm app}$ given above should be considered an upper limit to the
actual value and, therefore, the estimates given below should be
considered as limits. In particular the derived distance $R$ is un
upper limit, the density is a lower limit and thus the clumping
factor should be considered as a lower limit. In Fig. 4 we show a
few derived quantities as a function of the aperture angle $\theta$
and for $t_{\rm ill}=300$ and 1000 seconds.  The distance $R$ is of
the order of a few times $10^{15}$ cm; the density $n_{\rm e}$ is of
the order of $10^{14}$--$10^{15}$ cm$^{-3}$; $\Delta R/R$ is of the
order of a tenth, and the filling factor of the order of $f\sim
10^{-3}$.  The mass of the line emitting layer is $M_{\rm l}=
10^{-4}$--$10^{-2}$ $M_\odot$.  It is interesting to note that the
assumed (admittedly simple) geometry can also account for the line
duration, which yield an independent value of $\Delta R/R$ (dashed
line in the top panel of Fig. 4).  According to it, the illumination
angle turns out to be $\sim 45^\circ$ (i.e. where the dashed and solid
lines labeled as $\Delta R/R$ cross).

The portion $E_{\rm ill}$ of the total energy of the illuminating
continuum directly impinging on the reflecting matter will be

\begin{equation}
E_{\rm ill} \simeq {1 \over \eta} L_{\rm line} t_{\rm line} \simeq
3.5\times 10^{50} \,\,\, {\rm erg.} 
\end{equation}

The total apparent isotropic energy contained in the illuminating
component will be $E_{\rm iso}=E_{\rm ill} / \Delta \Omega$, where
$\Delta \Omega $ is the solid angle subtended by the reflecting matter
as observed the location of the illuminating source. It can be shown
that in the geometry assumed here and with the geometrical parameters
found above $\Delta \Omega \sim 0.01 \, 4\pi$, giving $E_{\rm
iso}\simeq 3\times 10^{51}$ erg, consistent with the early phase of
the afterglow.

We conclude that the reflection model offers a good interpretation of 
the data.  The reflecting material must be clumped, although the 
required clumping factor is smaller (and the corresponding filling factor 
greater) than the one required by the thermal 
model (Lazzati 2003). 
  
\begin{figure}   
\psfig{figure=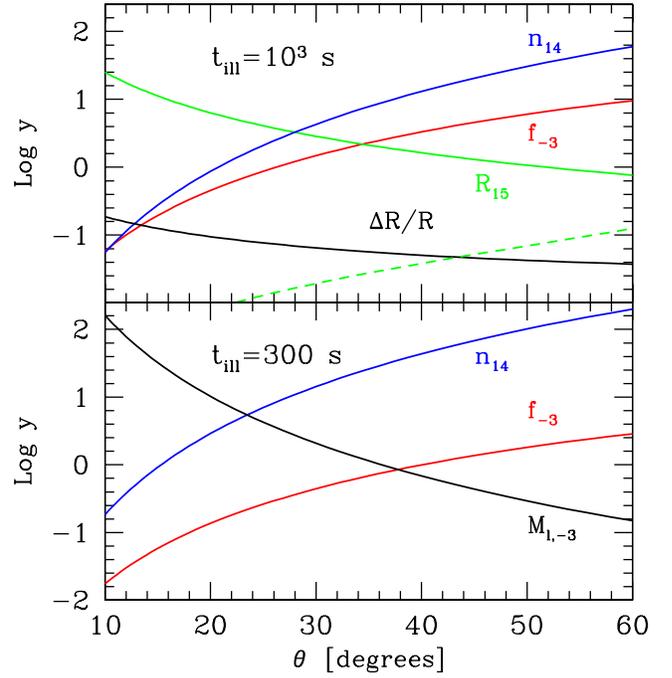,width=10cm}   
\vskip -0.7 true cm 
\caption{Derived quantities for GRB 021211, for $t_{\rm ill}=10^3$ s 
(top panel) and for $t_{\rm ill}=300$ s (bottom panel). The filling 
factor is in units of $10^{-3}$, the distance $R$ in units of 
$10^{15}$ cm, the density $n$ in units of $10^{14}$ cm$^{-3}$ and the 
mass of the emitting layer in units of $10^{-3}$ solar masses.  The 
quantities $R$, $\Delta R/R$ and $M_{\rm l}$ do not depend on $t_{\rm 
ill}$.  The dashed line in the top panel is the value of $\Delta R /R$ 
derived using Eq. (\ref{r}) and Eq. (\ref{dr}).  See text for the 
assumed values of the input parameters.} 
\label{parameters} 
\end{figure}

\subsection{Re--emission of absorbed radiation}   
   
An important consequence of the assumption that reflection is   
responsible for the production of detected lines is that, especially   
when $\xi $ is low ($\xi <10^2$) as in the case discussed here, most   
of the incoming radiation is absorbed by the reflecting material and   
successively {\it re--emitted} as low--frequency radiation.   
   
In the low--ionization condition discussed here the reflected 
luminosity is $\sim 10 \%$ of the total incident luminosity 
(e.g. Zicky et al. 1994): therefore the remaining $\sim 90\%$ is 
absorbed by the medium, which is heated--up and re--emits what it has 
absorbed.  The luminosity $L_{\rm repr}$ of this reprocessed component 
is rather important and amounts to: 
\begin{equation}   
L_{\rm repr}\, =\, 0.9\, {L_{\rm line} \over \eta}\, \sim  
9\times 10^{46} \, {L_{\rm line,45} \over \eta_{-2}} \quad {\rm erg\, s^{-1}}  
\label{lrepr} 
\end{equation}   
This emission lasts exactly for the same time as the X--ray emission 
line, and is therefore a powerful diagnostic of the reflection model. 
The above luminosity, for the case of GRB 011211 at $z=2.14$, 
corresponds to a flux of $3\times 10^{-12}$ erg s$^{-1}$ cm$^{-2}$. 
The problem is to assess the spectral region where this emission 
peaks.  We should solve the radiation transfer into a slab, heated by 
the incoming rapidly variable high energy emission from the early 
afterglow.  The surface of the slab, directly illuminated by the 
incoming continuum will be heated to large temperatures, close to the 
Compton temperature $T_C\sim 10^7$ K.  The emission from these outer 
portion will peak in the X--ray region, at energies around 1 keV.  Gas 
in the layers deeper into the slab will be close to the thermal 
equilibrium: therefore the temperature of these region will be close 
to the black--body temperature corresponding to a black--body emitting 
a luminosity of the order of $L_{\rm bb}=L_{\rm repr}$: 
\begin{equation}   
T_{\rm bb}\, \simeq\, \left( \frac {L_{\rm bb}}{\sigma A}\right)^{1/4}\,  
\simeq\,  
2\times 10^5 \left( {L_{\rm repr, 47}\over A_{30}} \right)^{1/4}   
\quad {\rm K} 
\label{tbb}
\end{equation}   
\noindent   
where the area $A=2\pi R\sin\theta\Delta R$ has been assumed  
of the order of $A=10^{30}A_{30}$ cm$^2$. 
The maximum of the emission falls at the frequency  
$\nu _{\rm bb}=3kT/[(1+z)h]\sim 10^{15}$ Hz, i.e. in the near UV. 
 
The actual outcoming spectrum will be a complex integral over the 
emission from the different layers (weighted by the optical depth of 
the layer), in strongly time--dependent conditions.  The detailed 
calculation of the spectrum is beyond the scope of this paper. 
Moreover, note that the conditions appropriate to analyze even a 
steady state case are at the limit of the calculation possibility of 
the program CLOUDY (Ferland 1996) which gives results which are not 
completely reliable. 
 
What we can safely conclude is that, simultaneously with the presence 
of the X--ray emission lines from light metals, there should be a 
bright component between the UV and the soft X--rays, which has a 
luminosity $\sim$100 times larger that the line luminosities (and 10 
times larger than the entire X--ray reflection continuum).  The low 
frequency tail of this reprocessed component falls in the optical 
band, where can be observed as a rebrightening of the light curve. 
As an illustrative example, if 1 per cent of the reprocessed 
luminosity is emitted in the optical V--band, we should see a 
magnitude of $V\sim 22$. 
 
We stress that the presence of the reprocessed emission is a natural 
consequence of the reflection interpretation: observations in the 
UV--soft X--ray bands, confirming or ruling out the existence of this 
component accompanying the X--ray emission lines, provide a powerful 
and reliable diagnostic tool to study the origin of the emission 
features. The only way we envisage not to see this reprocessed 
component in the reflection scenario (with a small ionization 
parameter) is to have dense and small clumps with a Thomson optical 
depth close to unity.  These blobs would efficiently reflect the 
incoming continuum, but only a small fraction of it is absorbed and 
re--emitted.  Since this requires a rather ad--hoc setup, we consider 
it unlikely. 

\subsection{Close reprocessor}

We discussed the case of a far reprocessor, located at a
distance $R\sim 10^{15}-10^{16}$ cm from the illuminator.  There is
however the possibility that the material producing the lines is
located very close to the central source. In this case the duration of
the line is linked to the time for wich the central engine is
active. A model of this type has been proposed by 
Rees \& M{\' e}sz{\' a}ros (2000) and M{\' e}sz{\' a}ros
\& Rees (2001).

The discussion given in the previous Sections can be easily applied to
this different scenario. In this case Eq. (\ref{r}) can be dropped,
since the geometrical effects are not important in determining the
time at which the lines appear. All the other equations still apply,
once $\theta $ is fixed to $\pi/2$. In this way we can model a
cilindrical geometry, with the illuminating source located on the
axis.  We fixed the size to $R=10^{13}$ cm, the typical size
considered by Rees \& M{\' e}sz{\' a}ros (2000) and M{\' e}sz{\' a}ros
\& Rees (2001).

We can derive the interesting physical parameters for the value of the
observational quantities given above in the same way as above. We
found: $f=4$, $n_e=1.2\times 10^{20}$ cm$^{-3}$, $\Delta R/R=0.03$ and
$M_l=2.5\times 10^{-8}$ M$_{\odot}$. The large value of the density is
a consequence of the fact that, for a fixed value of the ionization
parameter, Eq. (\ref{xidef}) implies a large density for compact
(i.e. small $R$) reprocessor. The value of $f$ (close to 1) shows that
no clumping is needed in this case, due to the smaller quantity of
iron required compared to the far reprocessor case.

Although the values of the derived physical quantities are reasonable,
this version of the reflection scenario suffers of many
problems. First of all (e.g. Lazzati et al. 2002b), the lines and the
continuum responsible for that lines are observed together: thus it is
not possible to produce lines with the large observed $EW$ (hundreds
of eV). This problem is naturally by-passed by the far-reprocessor
scenario, since, due to the crossing time delay, the lines are
observed much later than the direct continuum. A possibility to solve the
difficulty in the close reprocessor is to admit that the illuminating
continuum is strongly anisotropic and threfore it is not directly
observable.

Another problem is related to the large temperature reached by the
illuminated material. In fact, following the results of Sect. 3.3, the
temperature of the material will lie between $T_{\rm bb}$ and
$T_C$. Due to the small area, $A\simeq 10^{25}$ cm$^{2}$,
Eq. (\ref{tbb}) gives $T_{\rm bb}\sim 3.5\times 10^6$ K. The
reprocessed thermal radiation will therefore peak in the soft X-ray
domain ($> 0.1$ keV). A strong soft excess has never been observed in
the X-ray spectra of afterglows with lines.

For these reasons we consider this possibility unlikely.

\subsection{Time evolution}   
   
All the issues related to the GRB are characterized by a strong 
time--dependent phenomenology.  Besides the normal afterglow behavior 
produced by the decelerating fireball, in the X--ray band we have the 
possibility to see the reflection component, which it will too be 
variable in flux and possibly in shape.  The visibility of the 
reflected component with the associated emission lines depends upon 
the relative strength of the afterglow and of the reflected component, 
and on the ionization parameter determining which lines are most 
efficiently produced.  In general, the reflection component will be 
visible (if at all) only in intermediate phases of the afterglow: on 
one hand the luminosity of the afterglow must decrease 
sufficiently to allow the reflection to be observed, but on the other 
hand the reflected radiation can be observed only for a time smaller 
than the light travel time of the line emitting material (after that 
time we expect the reflection component to disappear). 
 
At these intermediate times, if the reflection component contributes 
to the total X--ray flux, it will produce a ``bump" in the X--ray 
lightcurve, flagging the likely presence of lines in the X--ray 
spectrum.  At the same time, we expect a change in the slope of the 
X--ray spectrum, which will depend upon the ionization parameter. In 
fact, for relatively small $\xi$, the reflecting layer will not be 
completely ionized, and the reflection component will be hard 
(i.e. characterized by a photon spectral index $\Gamma<2$) in the 
0.1--10 keV band, while for large $\xi$ the reflecting layer will be 
almost completely ionized, and the shape of the reflected component 
would closely resemble the shape of the illuminating continuum.  These 
simple arguments offer a direct test of the reflection scenario: 
suppose that the X--ray light curve, after a few hours, flattens: if 
this flattening is due to the emergence of the reflection component 
then its spectrum should be harder or almost equal to the spectrum of 
the ``normal'' afterglow.  If it is harder we expect a small ionization 
parameter and the presence of emission lines from light metals (and 
weak or absent lines from iron, cobalt or nickel).  If the spectrum 
maintains a constant or quasi--constant shape, then we expect the 
presence of K$_\alpha$ iron (or cobalt o nickel) lines and no emission 
from the fully ionized light metals, unless the ionization parameter 
is so large to maintain also the iron fully ionized. 
 
In conclusion we have presented a coherent scenario in which the 
emission of lines from light metals can be explained as due to the 
reflection by dense matter around the GRB. This scenario can be tested 
by observations. In particular the prediction of a bright optical-UV 
component coming together with the appearence of the emission lines 
offers a strong test to the overall picture. 

\begin{acknowledgements}   
We thank the referee for helpful criticisms that help us to improve the
paper.  We thank the Italian MIUR and ASI for financial support.
\end{acknowledgements}

\end{document}